\documentclass[12pt,aps,prl,floatfix,tightenlines]{revtex4}
\pdfoutput=1
\usepackage[pdftex]{hyperref}
\usepackage{graphicx}
\usepackage{color}
\usepackage{amsmath}
\usepackage{latexsym}

\newcommand{\ket}[1]{\vert#1\rangle}
\newcommand{\bra}[1]{\langle#1\vert}
\newcommand{\sqinv}[1]{\frac{1}{\sqrt{#1}}}
\newcommand{\rme}{\mathrm{e}}
\newcommand{\rmi}{\mathrm{i}}

\begin{document}

\title{Experimental violation of Svetlichny's inequality}
\author{J. Lavoie, R. Kaltenbaek, and K.J. Resch}
\email{kresch@iqc.ca}
\affiliation{Institute for Quantum Computing and Department of Physics \&
Astronomy, University of Waterloo, Waterloo, Canada, N2L 3G1}

\begin{abstract}
\noindent It is well known that quantum mechanics is incompatible
with local realistic theories. Svetlichny showed, through the
development of a Bell-like inequality, that quantum mechanics is
also incompatible with a restricted class of \emph{nonlocal}
realistic theories for three particles where any two-body nonlocal
correlations are allowed \cite{svetlichny87}.  In the present work,
we experimentally generate three-photon GHZ states to test
Svetlichny's inequality. Our states are fully characterized by
quantum state tomography using an overcomplete set of measurements
and have a fidelity of $(84\pm1)\%$ with the target state. We
measure a convincing, $3.6\sigma$, violation of Svetlichny's
inequality and rule out this class of restricted nonlocal realistic models.
\end{abstract}
\maketitle

\section{Introduction}
Quantum mechanics cannot be described by local hidden-variable (LHV)
theories. This is the conclusion of Bell's seminal work, in which he
derived a strict limit to the strength of correlations achievable by
all LHV models that is violated by quantum predictions
\cite{bell64}. Bell's original inequality did not allow for
imperfections and thus it was not accessible to experimental tests.
Clauser, Horne, Shimony and Holt (CHSH) addressed this issue and
developed the CHSH inequality~\cite{clauser69}, which allowed for tests in
actual experiments. Since then a growing number of experiments have
been reported (for examples,
see~\cite{freedman72,fry76,aspect82a,aspect82b,ou88,shih88,tapster94,kwiat95,
tittel98,weihs98,rowe01,aspelmeyer03,resch05a,ursin07,matsukevich08,
kaltenbaek08}),
and the overwhelming experimental evidence from these tests is in
favour of quantum mechanics, ruling out LHV theories.
It should be noted that, while no \emph{loophole-free} Bell test has been
performed, the most significant potential loopholes, relating to detection
efficiency and space-like separation of the choices of measurements settings,
have both separately been closed \cite{aspect82b,weihs98,rowe01}.

Both Bell's inequality and the CHSH inequality were formulated for
testing the correlations between just two particles. For more than
two particles, Greenberger, Horne and Zeilinger (GHZ)
showed~\cite{greenberger89} that a contradiction between LHV
theories and quantum mechanics can be seen directly in perfect
correlations, as opposed to statistically in imperfect ones. Soon
thereafter Bell-type inequalities for more than two particles were
developed \cite{mermin90, ardehali92, klyshko93, belinskii93,
zukowski97, gisin98, cabello02, guhne05, scarani05}. Quantum
predictions can violate such inequalities by an amount increasing
exponentially with the particle number \cite{mermin90, ardehali92,
belinskii93, zukowski97, gisin98, guhne05}.

All of the aforementioned inequalities are based on the assumption
that local realism applies to each individual particle. Two-particle
inequalities have been developed which are in conflict with quantum
mechanics although they allow restricted, but physically motivated,
nonlocal correlations \cite{leggett03}.  These inequalities have
recently been violated experimentally \cite{groeblacher07}.

Svetlichny showed that even if one allows unrestricted nonlocal
correlations between any two of the constituent particles in a
three-particle setting one can still find inequalities violated by
quantum mechanical predictions \cite{svetlichny87}. The correlations
allowed by Svetlichny's model are strong enough to maximally violate 
three-partite inequalities, such as Mermin's \cite{cereceda02}, which assume 
local realism for all particles involved. A violation of such inequalities
therefore can only rule out LHV theories, while a violation of
Svetlichny's inequality directly rules out a whole class of nonlocal
hidden-variable theories \cite{seevinck01,cereceda02,mitchell04,toth05,ghose08}.
Svetlichny's work has since been generalized to the case of $N$
particles \cite{seevinck02, collins02,laskowski05}.

Experimental tests have been performed confirming the violation of
the Mermin inequality \cite{pan00}, the
Mermin-Ardehali-Belinskii-Klyshko (MABK) inequality \cite{zhao03},
and the cluster state inequality developed by Scarani et. al.
\cite{walther05a}. For an even number of particles only, a
sufficiently large violation of the MABK inequality also rules out
partially non-local hidden-variable models \cite{collins02}. The
violation of the MABK inequality in \cite{zhao03} thus confirmed
genuine four-particle entanglement and non-locality. The original
Svetlichny inequality, however, remains untested.

Here, we begin with a brief theoretical description of Svetlichny's
inequality. We then experimentally produce high-fidelity
three-photon GHZ states and characterize them via quantum state
tomography \cite{resch05b}. Using these states, modulo standard loopholes
\cite{aspect82b,weihs98,rowe01}, we experimentally demonstrate a
convincing violation of Svetlichny's inequality.

\section{Theory}
The two assumptions resulting in Bell-type inequalities are locality
and realism as they were introduced by Einstein, Podolsky and Rosen
\cite{einstein1935}. We first review a straightforward method
to derive the CHSH inequality from these two assumptions following
an argument described by Peres \cite{peres93}. Pairs of particles
are distributed to two distant parties, A and B. Party A (B) can
choose between two measurement settings $\mathbf{a}$ and
$\mathbf{a^\prime}$ ($\mathbf{b}$ and $\mathbf{b^\prime}$). For each
measurement setting, two outcomes, $+1$ or $-1$, are possible.
Realism assumes that the measurement outcomes are predetermined by
some properties of the system investigated. These properties are known as
hidden variables because they are not necessarily accessible to
observation. The additional assumption of locality requires that the
measurement outcomes on side A are independent of the measurement
setting on side B, and vice versa. Thus for any given pair of
particles the measurement outcomes have predetermined values $a =
\pm 1$ and $a^\prime = \pm 1$ on side A and $b = \pm 1$ and
$b^\prime = \pm 1$ on side B. These values identically satisfy the
relations:
\begin{eqnarray}
S_2 & \equiv & a (b + b^\prime) + a^\prime (b - b^\prime) = \pm 2 \nonumber\\
S^\prime_2 & \equiv & a^\prime (b^\prime + b) + a (b^\prime - b) =
\pm 2. \label{S2}
\end{eqnarray}
If, for example, $S_2$ is averaged over many trials, the absolute
value must be smaller than $2$, which results in the CHSH inequality
\cite{clauser69}:
\begin{equation}
\left| E(\mathbf{a},\mathbf{b}) + E(\mathbf{a},\mathbf{b^\prime}) +
E(\mathbf{a^\prime},\mathbf{b}) -
E(\mathbf{a^\prime},\mathbf{b^\prime})\right| \le 2,
\end{equation}
where the correlation, $E(\mathbf{a},\mathbf{b})$, is the ensemble
average $\langle a b\rangle$ over the product of measurement
outcomes $a$ and $b$ for measurement settings $\mathbf{a}$ and
$\mathbf{b}$, respectively.

This argument can be extended to three particles \cite{collins02}. We will
denote the particles as well as the measurement outcomes as $a$, $b$ and $c$,
the measurement settings as $\mathbf{a}$, $\mathbf{b}$, $\mathbf{c}$. The
outcome of each measurement can be $+1$ or $-1$. If we assume local
realism for each of the three particles, then
for a given set of three particles the measurement outcomes $a$, $b$ and $c$ as
well as their primed counterparts will have predetermined values $\pm 1$. Using
(\ref{S2}) we find that the following identity must hold:
\begin{equation}
S_3 \equiv S_2 (c + c^\prime) + S^\prime_2 (c - c^\prime) =
2 \left(a^\prime b c + a b^\prime c + a b c^\prime -
a^\prime b^\prime c^\prime\right)
= \pm 4\label{S3}.
\end{equation}
Dividing this expression by two, and averaging over many trials yields
Mermin's inequality \cite{mermin90} for three particles:
\begin{equation}
\left|E(\mathbf{a^\prime},\mathbf{b},\mathbf{c}) +
E(\mathbf{a},\mathbf{b^\prime},\mathbf{c}) +
E(\mathbf{a},\mathbf{b},\mathbf{c^\prime}) -
E(\mathbf{a^\prime},\mathbf{b^\prime},\mathbf{c^\prime})\right| \le
2\label{Mermin},
\end{equation}
where $E(\mathbf{a},\mathbf{b},\mathbf{c}) = \langle a b c\rangle$.

Now assume that we allow arbitrary (nonlocal) correlations between just
\emph{two} of the particles, say $a$ and $b$, while we still assume local
realism with respect
to the third particle, $c$. In this case we cannot factorize $S_2$ as we did in
(\ref{S2}) because outcomes for particle $a$ might nonlocally depend on the
outcomes and/or measurement settings for particle $b$. However, we can
still write
\begin{eqnarray}
\tilde{S}_2 &=&(a b) + (a b^\prime) + (a^\prime b) - (a^\prime
b^\prime)\nonumber \\
\tilde{S}_2^\prime &=& (a^\prime b^\prime) + (a^\prime b) + (a
b^\prime) - (a b),
\end{eqnarray}
where the parentheses are meant as a reminder that these quantities should be
regarded as separate and independent quantities. Each of these quantities as
well as $c$ and $c^\prime$ must take predetermined values $\pm 1$ because we
assume local realism with respect to the third particle.
This model is strong enough to violate, and reach the algebraic maximum of
Mermin's inequality (since $\tilde{S}_2$ can be $\pm 4$). Thus no experimental 
violation of Mermin's inequality can rule out this restricted nonlocal 
hidden-variable model.

With this in mind let us slightly modify our argument to derive Svetlichny's
inequality. Because $\tilde{S}_2$ and $\tilde{S}^\prime_2$ are functions 
of the same four quantities, they are not independent. For example, whenever 
one of the two quantities reaches its algebraic maximum
$\pm 4$, the other one will be $0$. As a result the following identity
holds:
\begin{flushleft}
\begin{eqnarray}
\tilde{S}_2 c - \tilde{S}^\prime_2 c^\prime & = &
\left(a b\right) c + \left(a b\right) c^\prime + \left(a
b^\prime\right) c - \left(a b^\prime\right) c^\prime +
\left(a^\prime b\right) c - \left(a^\prime b\right) c^\prime -
\left(a^\prime b^\prime\right) c -
\left(a^\prime b^\prime\right) c^\prime \nonumber\\
& = & \pm 4, \pm 2, 0\label{SValgebraic}.
\end{eqnarray}
\end{flushleft}
Averaging over many trials yields the Svetlichny inequality:
\begin{eqnarray}
\mathcal{S}_v & \equiv  &\left|
E(\mathbf{a},\mathbf{b},\mathbf{c}) +
E(\mathbf{a},\mathbf{b},\mathbf{c^\prime}) +
E(\mathbf{a},\mathbf{b^\prime},\mathbf{c}) -
E(\mathbf{a},\mathbf{b^\prime},\mathbf{c^\prime}) + \right. \nonumber\\
& &\left. E(\mathbf{a^\prime},\mathbf{b},\mathbf{c}) -
E(\mathbf{a^\prime},\mathbf{b},\mathbf{c^\prime}) -
E(\mathbf{a^\prime},\mathbf{b^\prime},\mathbf{c}) -
E(\mathbf{a^\prime},\mathbf{b^\prime},\mathbf{c^\prime})
\right| \le 4\label{Svetlichny},
\end{eqnarray}
where we refer to $\mathcal{S}_v$ as the Svetlichny parameter.
It is remarkable that, although we started out by allowing nonlocal
correlations between particles $a$ and $b$ while $c$ is local, one
gets an expression identical to (\ref{Svetlichny}) if $b$ and $c$
are nonlocally correlated while $a$ is local, or if $a$ and $c$ are
nonlocally correlated while $b$ is local. Every hidden-variable
model that allows for nonlocal correlations between any two
particles but not between all three can be seen as a probabilistic
combination of models where the partition of the particles between
nonlocal and local is made one or the other way. All of these models
fulfill the Svetlichny inequality \cite{mitchell04}.

It was shown by Svetlichny that his inequality can be violated by
quantum predictions, and that the maximum violation can be achieved
with GHZ states.  Assume we have a polarization-entangled GHZ state
$\ket{\psi} = \sqinv{2} \left(\ket{HHV}+\ket{VVH}\right)$, and let
our measurement settings all be in the \textit{xy}-plane of the
Bloch sphere, i.e.~we can write the corresponding states we project on
as $\sqinv{2}\left(\ket{H} +
\rme^{\rmi \phi} \ket{V}\right)$. For example, the measurement settings
$\mathbf{a}$ and $\mathbf{a^\prime}$ for particle $a$ correspond to
projective measurements on the states 
$\ket{A(\pm)} = \sqinv{2}\left(\ket{H} \pm \rme^{\rmi
\phi_a}\ket{V}\right)$ and $\ket{A^\prime(\pm)} =
\sqinv{2}\left(\ket{H} \pm \rme^{\rmi \phi^\prime_a}\ket{V}\right)$,
respectively. Here, the $\pm$ corresponds to the state the particle
is projected on if the outcome of the measurement is $\pm 1$. For
particles $b$ and $c$ we choose an analogous notation. Then the
quantum prediction for the left-hand side of (\ref{Svetlichny}) is
\begin{eqnarray}
& & \left| \cos\left(\phi_a + \phi_b - \phi_c\right) +
\cos\left(\phi_a + \phi_b - \phi_c^\prime\right) +
\cos\left(\phi_a + \phi_b^\prime - \phi_c\right) -
\right.\nonumber\\
& & \left.
\cos\left(\phi_a + \phi_b^\prime - \phi_c^\prime\right) +
\cos\left(\phi_a^\prime + \phi_b - \phi_c\right) -
\cos\left(\phi_a^\prime + \phi_b - \phi_c^\prime\right) -
\right.\nonumber\\
& & \left.
\cos\left(\phi_a^\prime + \phi_b^\prime - \phi_c\right) -
\cos\left(\phi_a^\prime + \phi_b^\prime - \phi_c^\prime\right) \right|.
\end{eqnarray}
With a suitable choice of angles, such as:
\begin{equation}
\phi_a = \frac{3\pi}{4},
\phi_a^\prime = \frac{\pi}{4},
\phi_b = \frac{\pi}{2},
\phi_b^\prime = 0,
\phi_c = 0,
\phi_c^\prime = \frac{\pi}{2}\label{angles},
\end{equation}
this results in $\mathcal{S}_v = 4\sqrt{2}\not\le 4$, which is the
maximum violation of Svetlichny's inequality achievable with
quantum mechanics \cite{mitchell04}.

Since any hidden-variable model describing a three-particle state where 
only two particles are nonlocally correlated has to fulfill the Svetlichny 
inequality, its violation explicitly rules out this type of nonlocal 
hidden-variable theory.

\begin{figure}
\begin{center}
\includegraphics[width=0.9 \columnwidth]{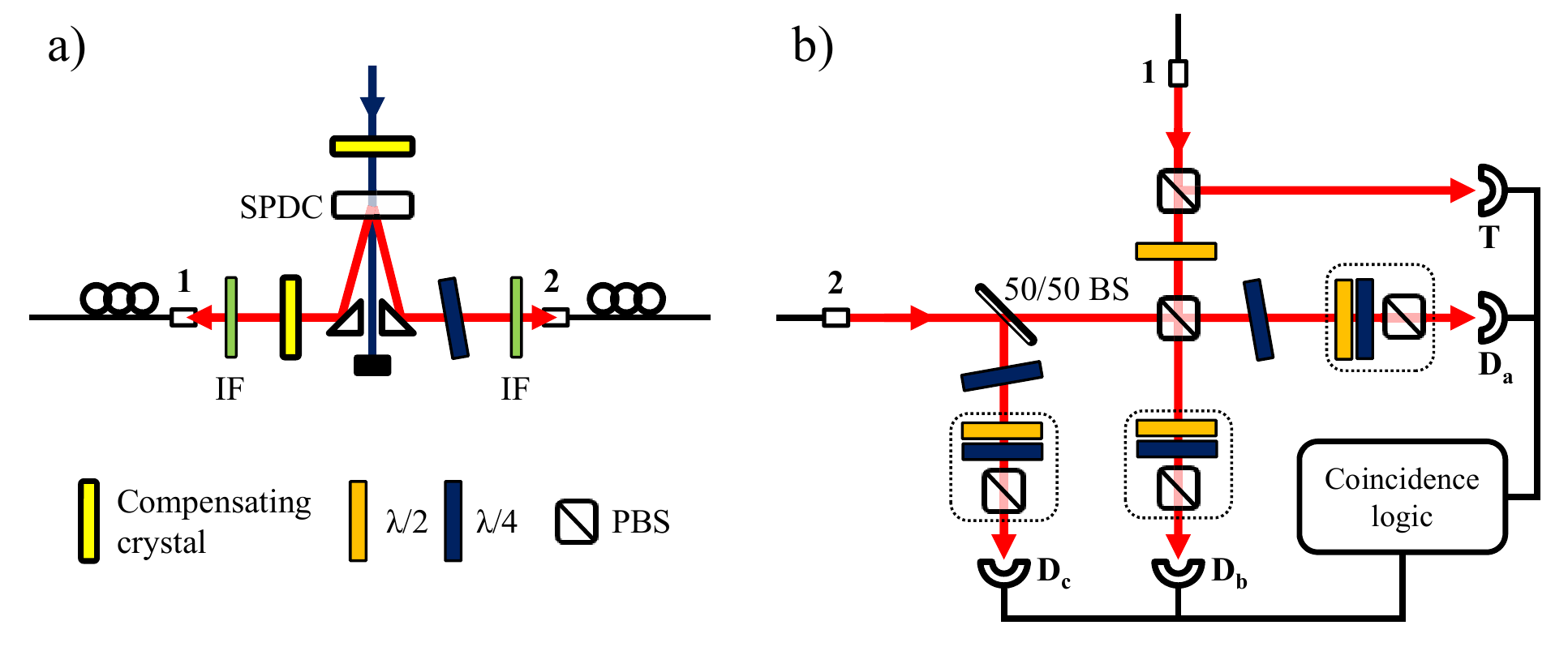}
\caption{Experimental setup. (a) Schematic of our type-I SPDC
source. A $45^\circ$ polarized, pulsed UV beam pumps a pair of
orthogonally oriented BBO crystals cut for type-I phase matching.
Temporal walk-off between the pairs created in the first and in the
second crystal is compensated by a 1~mm thick $\alpha$-BBO and two
quartz crystals (one 2~mm and one 0.5~mm thick) before the SPDC
crystals. Spatial walk-off, which occurred in mode 1, was
compensated by a BiBO crystal. The phase between horizontal and
vertical photons was adjusted by tilting a $\lambda/4$ plate in mode
2. All photons pass through $3\,\mbox{nm}$ FWHM bandwidth filters
around $790\,\mbox{nm}$ and are coupled into single-mode
fibres corresponding to the spatial modes 1 and 2. (b) The
interferometer used to project on a three-photon GHZ state. Using
fibre polarization controllers, the polarization in mode 2 is
rotated such that we map $H \to V$ and $V \to H$, while in mode 1,
$H$ and $V$ are preserved. Inside the interferometer the four
photons from a double-pair emission can be split up into four
separate spatial modes and result in a four-fold coincidence event
between the detectors $T$, $D_a$, $D_b$, and $D_c$. In this case the
three photons in the modes $a$, $b$, and $c$ will be projected on
the three-photon GHZ state $\sqinv{2}\left(\ket{H_aH_bV_c} +
\ket{V_aV_bH_c}\right)$ given that the photons detected by $D_a$ and $D_b$
are indistinguishable. \label{setup}}
\end{center}
\end{figure}

\section{Experiment}

Our experiment uses a pulsed titanium:sapphire laser (rep. rate
80MHz, $2.5\,\mbox{W}$ avg. power, $790\,$nm center wavelength,
$9\,$nm FWHM bandwidth). We frequency double the near-infrared beam,
producing $700\,$mW average power near $395\,\mbox{nm}$ with a FWHM
bandwidth of $1.8\,$nm. This upconverted beam is focused on a pair
of orthogonally-oriented $\beta$-Barium-Borate (BBO) nonlinear
crystals~\cite{kwiat99} cut for type-I noncollinear degenerate
spontaneous parametric down-conversion (SPDC) with an external half
opening angle of $3^\circ$. The pump polarization is set to $\ket{D}
= \sqinv{2}(\ket{H}+\ket{V})$ such that each pump photon can produce
a photon pair either in the first or the second BBO crystal. To
compensate for temporal distinguishablity between the pairs created
in the first and the second crystal the pump passes through a $1$mm
$\alpha$-BBO crystal, a $2$mm quartz crystal, and a $0.5$mm quartz
crystal, all cut for maximum birefringence. To compensate the
$75\mu$m spatial walk-off between the horizontally and vertically
polarized SPDC photons observed in one of the output modes, we insert
a $0.75$mm thick BiBO crystal cut at $\theta = 152.6^\circ$ and
$\phi = 0^\circ$. For these cut angles the crystal compensates the
transverse walk-off without introducing additional time walk-off.
The photons are subsequently coupled into single-mode fibres.  We
label the two corresponding spatial output modes as 1 and 2, see
figure~1(a). Fiber polarization controllers ensure that in mode 1
states in the HV basis remain unchanged while in mode 2 we flip the
polarization, i.e., $H \leftrightarrow V$. With this configuration
we achieve a two-photon coincidence rate of $43\,\mbox{kHz}$ and
single rates of about $240\,\mbox{kHz}$ and $270\,\mbox{kHz}$ for
modes 1 and 2, respectively. The measured contrast of the pairs is
$75:1$ in the H/V basis and $61:1$ in the $45^\circ/ -45^\circ$ basis when the
source is adjusted to produce $\vert\phi^+\rangle = \sqinv{2} \left(
\vert HH\rangle + \vert VV\rangle\right)$ states.

Following the approach in \cite{bouwmeester99} we use the double-pair
emission of the SPDC source to produce 3-photon GHZ correlations in the
interferometer shown in figure~1(b). A four-fold coincidence detection
in the four outputs of the interferometer indicates the successful
generation of the GHZ state.
To lowest significant order, a four-fold coincidence can only
occur if two photons enter the interferometer via the spatial mode 1, and
their two partner photons enter it via mode 2.
The two photons in input mode 1 impinge on a polarizing beam splitter (PBS).
In order for a four-fold coincidence event to occur one of them must be H
polarized and the other V polarized. The V photon is reflected at the
PBS. Its detection by detector $T$ serves as a trigger event. The H photon
passes through the PBS in mode 1, and a $\lambda/2$ plate oriented
at $22^\circ$ rotates the polarization from $\ket{H}$ to $\ket{D}$. The two
photons in mode 2 are split at the $50:50$ beam splitter (BS) with
probability $1/2$. Only in this case a four-fold coincidence can occur. The
transmitted photon and the $\ket{D}$ polarized
photon in mode 1 are overlapped on a PBS. A coincidence detection event in
the two output modes of the PBS can only occur if both photons are
transmitted or both are reflected. If these two possibilities are
indistinguishable, Hong-Ou-Mandel interference will occur \cite{hong1987}.
In the reflected mode of the BS we compensate for a phase shift due
to birefringence in the BS by tilting a $\lambda/4$ plate.

A four-fold coincidence detection in the interferometer
outputs can only occur if the trigger photon is $V$ polarized, and if the
other three photons in the modes $a$, $b$ and $c$ (see figure 1(b)) are
either $\ket{H_aH_bV_c}$ or $\ket{V_aV_bH_c}$. By tilting a
$\lambda/4$ plate in output mode $a$ we adjust the relative phase between
these contributions such that a four-fold coincidence event signals a GHZ
state of the form
$\sqinv{2}\left(\vert H_aH_bV_c\rangle + \vert V_aV_bH_c\rangle\right)$.

Given this state, quantum mechanics predicts a maximum violation of
Svetlichny's inequality if we choose measurements of the form
$\ket{H}+\rme^{\rmi\phi}\ket{V}$ with the angles given in (\ref{angles}).
Particles $a$, $b$ and $c$ are identified with
photons in the interferometer output modes $a$, $b$ and $c$, respectively.
To test Svetlichny's inequality each of the photons has to
be measured in two measurement bases, and for each basis there are
two possible outcomes, $+1$ and $-1$. For each outcome we have to
set the polarization analyzer in the respective mode such that a photon
detected after passing through the analyzer corresponds to that outcome.

\begin{figure}
\begin{center}
\includegraphics[width=0.4 \columnwidth]{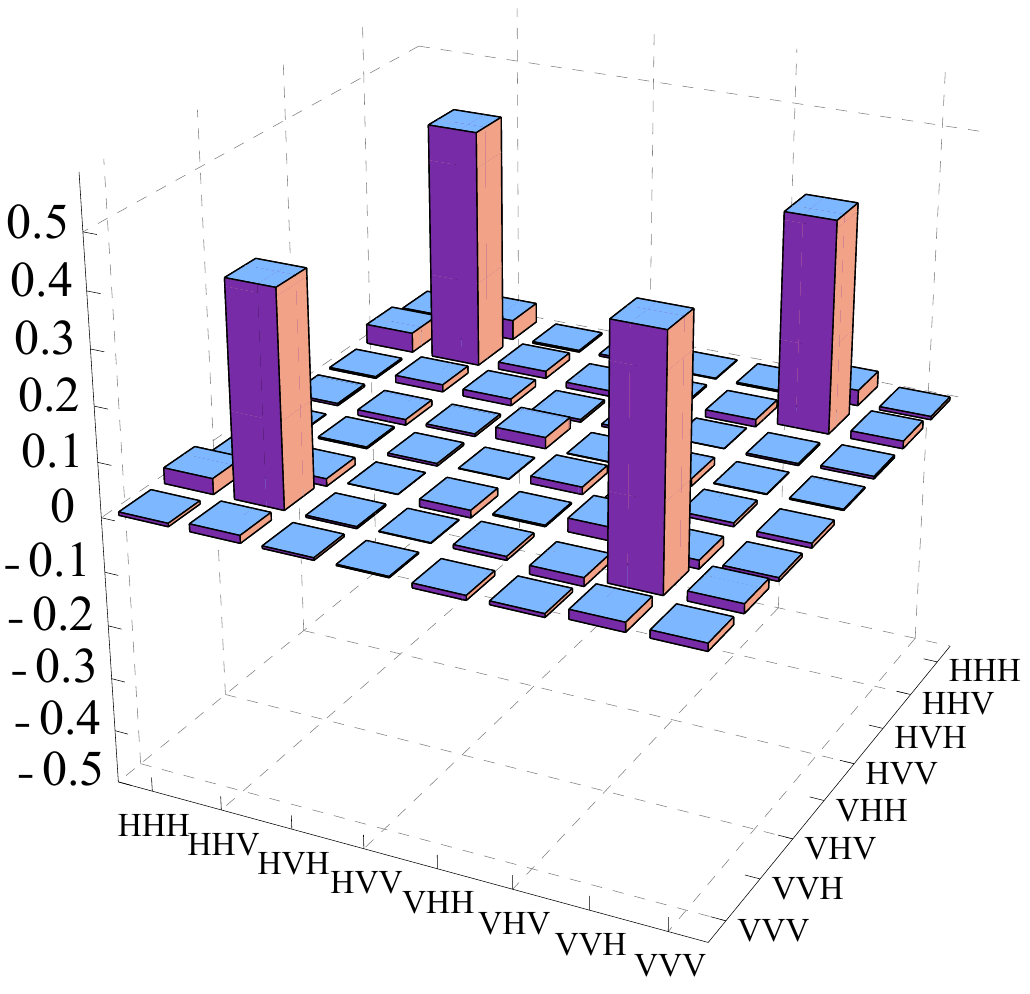}
\includegraphics[width=0.4 \columnwidth]{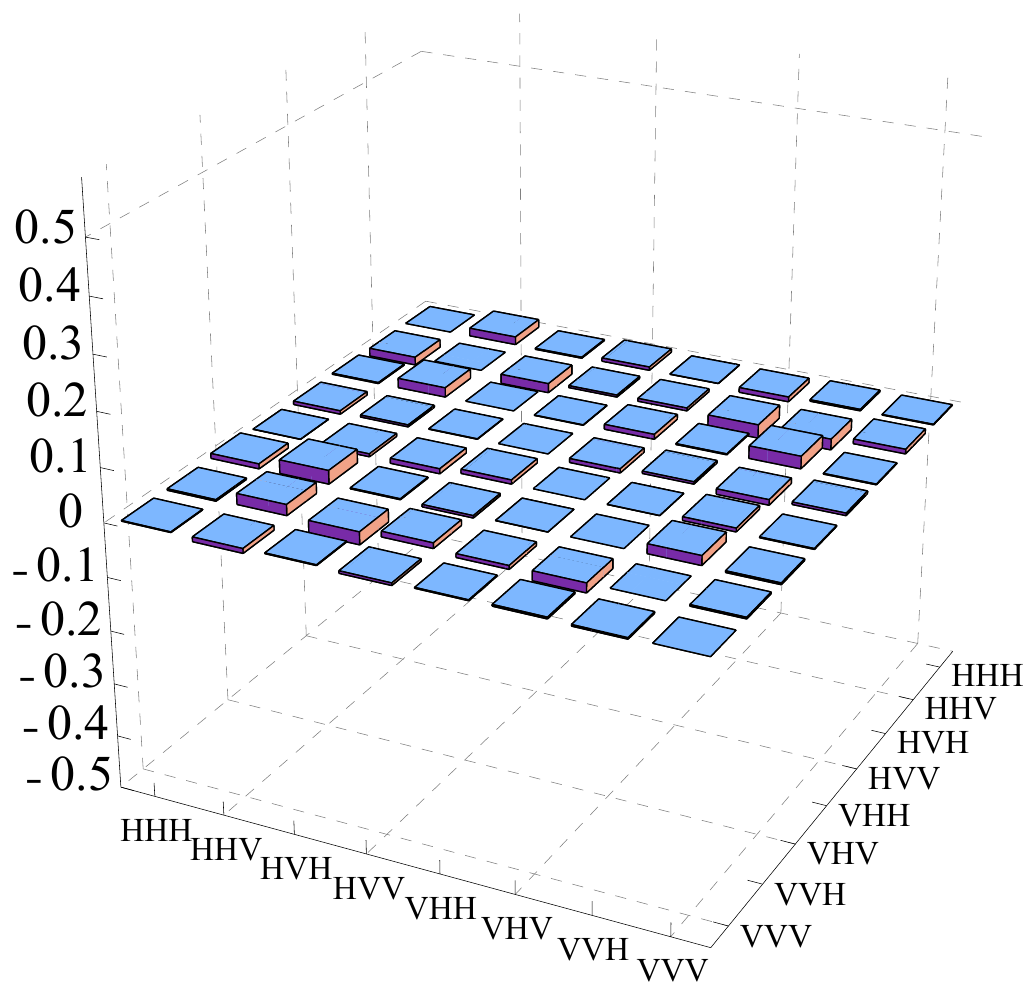}
\caption{Reconstructed three-photon density matrix.  a) Real part
and b) imaginary part of the density matrix. The state was
reconstructed from a tomographically overcomplete set of $216$
measurements. For each measurement the four-fold coincidence counts
were integrated over $27\,\mbox{min}$, see table \ref{counts}. The
fidelity of the reconstructed density matrix with the GHZ state
$\frac{1}{\sqrt{2}}(|HHV\rangle+|VVH\rangle)$ is $(84\pm1)\%$.
\label{dm}}
\end{center}
\end{figure}

The Svetlichny parameter, $\mathcal{S}_v$ (see (\ref{Svetlichny})), consists 
of $8$ correlations, each of which can
be constructed from $8$ three-photon polarization measurements for a
total of $64$ measurements. Each polarization analyzer consists of a
$\lambda/2$-plate followed by a $\lambda/4$-plate and then a PBS.
The photons passing the PBSs are detected with single-mode
fibre-coupled single-photon counting modules; the coincidence window
is $10\,\mbox{ns}$.

In order to fully characterize the state produced by our setup we
perform quantum state tomography. Because all the
measurement settings for the Svetlichny inequality lie in the
\textit{xy}-plane of the Bloch sphere these alone are not
tomographically complete. Instead of performing an additional run
apart from the measurements of the Svetlichny settings, we add
two additional projective measurements ($\ket{H}$ and $\ket{V}$) for each of 
the particles, resulting in a total of $216$ three-photon
polarization measurements. Our set of measurements is now
actually tomographically overcomplete, which has been shown to produce 
better estimates of quantum states \cite{deburgh08}.

\section{Results}

\begin{table}
 \begin{center}
   {\tiny
  \begin{tabular}{|c|c||c|c|c|c|c|c|}
   \hline
   \multicolumn{2}{|c||}{Settings for} & \multicolumn{6}{|c|}{Settings for a} \\
   \cline{3-8}
   b & c & $\ket{A(+)}$ &
   $\ket{A(-)}$ &
   $\ket{A^\prime(+)}$ &
   $\ket{A^\prime(-)}$ &
   $\ket{H}$ &
   $\ket{V}$ \\
   \hline
$\ket{B(+)}$ & $\ket{C(+)}$ & $\textbf{12}$ & $\textbf{56}$ & $\textbf{20}$ & $\textbf{69}$ & $53$ & $35$\\
 & $\ket{C(-)}$ & $\textbf{52}$ & $\textbf{19}$ & $\textbf{47}$ & $\textbf{18}$ & $38$ & $32$\\
 & $\ket{C^\prime(+)}$ & $\textbf{12}$ & $\textbf{50}$ & $\textbf{43}$ & $\textbf{13}$ & $31$ & $35$\\
 & $\ket{C^\prime(-)}$ & $\textbf{48}$ & $\textbf{16}$ & $\textbf{14}$ & $\textbf{56}$ & $31$ & $23$\\
 & $\ket{H}$ & $34$ & $39$ & $32$ & $44$ & $4$ & $71$\\
 & $\ket{V}$ & $35$ & $26$ & $30$ & $29$ & $62$ & $8$\\
   \hline
$\ket{B(-)}$ & $\ket{C(+)}$ & $\textbf{70}$ & $\textbf{16}$ & $\textbf{76}$ & $\textbf{12}$ & $36$ & $35$\\
 & $\ket{C(-)}$ & $\textbf{12}$ & $\textbf{53}$ & $\textbf{12}$ & $\textbf{46}$ & $39$ & $33$\\
 & $\ket{C^\prime(+)}$ & $\textbf{49}$ & $\textbf{8}$ & $\textbf{17}$ & $\textbf{59}$ & $37$ & $44$\\
 & $\ket{C^\prime(-)}$ & $\textbf{22}$ & $\textbf{40}$ & $\textbf{75}$ & $\textbf{19}$ & $39$ & $30$\\
 & $\ket{H}$ & $47$ & $37$ & $28$ & $33$ & $4$ & $75$\\
 & $\ket{V}$ & $32$ & $24$ & $47$ & $34$ & $54$ & $4$\\
   \hline
$\ket{B^\prime(+)}$ & $\ket{C(+)}$ & $\textbf{19}$ & $\textbf{69}$ & $\textbf{57}$ & $\textbf{16}$ & $40$ & $31$\\
 & $\ket{C(-)}$ & $\textbf{51}$ & $\textbf{17}$ & $\textbf{6}$ & $\textbf{63}$ & $25$ & $40$\\
 & $\ket{C^\prime(+)}$ & $\textbf{48}$ & $\textbf{13}$ & $\textbf{47}$ & $\textbf{12}$ & $26$ & $37$\\
 & $\ket{C^\prime(-)}$ & $\textbf{15}$ & $\textbf{62}$ & $\textbf{18}$ & $\textbf{56}$ & $34$ & $28$\\
 & $\ket{H}$ & $32$ & $38$ & $34$ & $39$ & $4$ & $68$\\
 & $\ket{V}$ & $44$ & $34$ & $36$ & $40$ & $53$ & $4$\\
   \hline
$\ket{B^\prime(-)}$ & $\ket{C(+)}$ & $\textbf{68}$ & $\textbf{18}$ & $\textbf{17}$ & $\textbf{62}$ & $34$ & $36$\\
 & $\ket{C(-)}$ & $\textbf{19}$ & $\textbf{54}$ & $\textbf{45}$ & $\textbf{11}$ & $33$ & $29$\\
 & $\ket{C^\prime(+)}$ & $\textbf{18}$ & $\textbf{48}$ & $\textbf{17}$ & $\textbf{54}$ & $42$ & $33$\\
 & $\ket{C^\prime(-)}$ & $\textbf{55}$ & $\textbf{13}$ & $\textbf{62}$ & $\textbf{29}$ & $32$ & $31$\\
 & $\ket{H}$ & $26$ & $25$ & $39$ & $42$ & $1$ & $51$\\
 & $\ket{V}$ & $47$ & $35$ & $44$ & $29$ & $63$ & $10$\\
   \hline
$\ket{H}$ & $\ket{C(+)}$ & $42$ & $40$ & $39$ & $52$ & $77$ & $6$\\
 & $\ket{C(-)}$ & $31$ & $22$ & $31$ & $32$ & $52$ & $4$\\
 & $\ket{C^\prime(+)}$ & $40$ & $37$ & $38$ & $35$ & $53$ & $1$\\
 & $\ket{C^\prime(-)}$ & $41$ & $30$ & $36$ & $38$ & $65$ & $5$\\
 & $\ket{H}$ & $2$ & $3$ & $5$ & $5$ & $8$ & $5$\\
 & $\ket{V}$ & $79$ & $66$ & $72$ & $67$ & $119$ & $5$\\
   \hline
$\ket{V}$ & $\ket{C(+)}$ & $39$ & $46$ & $32$ & $43$ & $3$ & $72$\\
 & $\ket{C(-)}$ & $35$ & $36$ & $29$ & $39$ & $2$ & $44$\\
 & $\ket{C^\prime(+)}$ & $25$ & $42$ & $33$ & $43$ & $7$ & $59$\\
 & $\ket{C^\prime(-)}$ & $32$ & $43$ & $30$ & $31$ & $6$ & $62$\\
 & $\ket{H}$ & $62$ & $68$ & $54$ & $69$ & $3$ & $131$\\
 & $\ket{V}$ & $4$ & $7$ & $3$ & $6$ & $3$ & $1$\\
   \hline
  \end{tabular}}
 \end{center}
 \caption{\label{counts}Experimentally-measured counts. Four-fold coincidences 
for the $216$ measurements performed for quantum state tomography. A subset 
of $64$ measured counts were used to test Svetlichny's inequality; these 
counts are shown in boldface type.  We cycled through all of the measurements
$27$ times, counting for $60\,\mbox{s}$ for each measurement. The four-fold
coincidences given are the result of integrating over all of these
cycles.}
\end{table}

In our setup the maximum four-fold coincidence rates of
$7\times 10^{-2}\,\mbox{Hz}$ and $8\times 10^{-2}\,\mbox{Hz}$ were achieved
for the correlations $|H_aH_bV_c\rangle$ and $|V_aV_bH_c\rangle$, 
respectively. To get statistically significant counts we integrated over
$27\,\mbox{min}$ per measurement. In order to reduce the negative effects of
misalignment of the setup over time we realized this
integration by counting for $60\,\mbox{s}$ for each of the $216$ measurements,
then repeating the full cycle of measurements $27$ times. The resulting
counts are given in table \ref{counts}.

We applied the maximum likelihood technique \cite{james01} to
reconstruct the density matrix of our state. Its real and
the imaginary part are shown in
figure 2. The fidelity $F = \bra{\psi}\rho\ket{\psi}$ of the density matrix
with the GHZ state $\sqinv{2} \left(\ket{HHV}+\ket{VVH}\right)$ is
$0.84 \pm 0.01$. The errors of quantities derived from the reconstructed
density matrix were calculated via a Monte Carlo simulation, where we
used each of the measured counts as the mean of a Poissonian distribution.
According to these distributions we generated random counts and ran the maximum
likelihood algorithm. This procedure was repeated $400$ times, and we report 
the standard deviation and mean for quantities derived from these 
reconstructed states.

The $64$ measurements that quantum mechanics predicts to
violate Svetlichny inequality are among the $216$ measured.
After integrating over all $27$ cycles we get a Svetlichny parameter of
$\mathcal{S}_v = 4.51\pm 0.14$; the eight measured correlations are shown 
in figure 3. This value violates the Svetlichny inequality by $3.6$ standard
deviations. It is, however, in good agreement with the value,
$\mathcal{S}^{QM}_v = 4.48 \pm 0.11$, predicted by quantum mechanics
given the reconstructed density matrix.

\section{Conclusion}
We used the double-pair emission from a pulsed type-I SPDC source
and projected the photons onto a GHZ state using a linear optical
interferometer.  We fully characterized the generated state and
reconstructed the density matrix applying the maximum likelihood
technique \cite{james01} using an overcomplete set of measurements.
From the reconstructed density matrix, we found that our state
matched the target GHZ state with a fidelity of $(84\pm 1)\%$.

We experimentally demonstrated the violation of the original
Svetlichny inequality for a three-particle GHZ state with a value of
$4.51\pm0.14$, which is greater than $4$ by $3.6$ standard deviations. This 
value is in good agreement with that predicted by quantum mechanics from 
our reconstructed density matrix, $4.48\pm0.11$. By violating Svetlichny's
long-standing inequality, we have shown that the correlations exhibited by
three particles cannot be described by hidden-variable theories with at most
two-particle nonlocality.

\begin{figure}
\begin{center}
\includegraphics[width=0.9 \columnwidth]{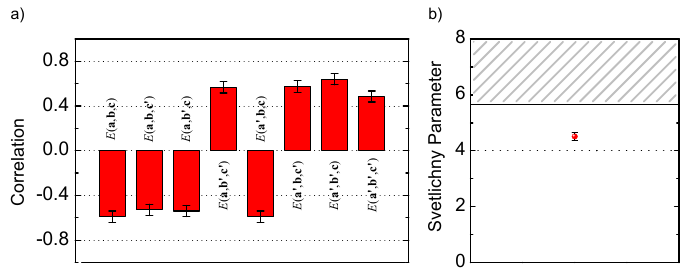}
\caption{Measured correlations and Svetlichny parameter. a) Measured
correlations for the eight combinations of measurement settings for
the three particles. Each correlation is constructed from $8$
four-fold coincidence measurements integrated over $27\,\mbox{min}$. 
The count rates for each of these measurements are given in 
table \ref{counts}. b) These
correlations yield a Svetlichny parameter of $4.51\pm 0.14$, which
clearly violates the bound (dashed line) of $4$ of the Svetlichny
inequality. The quantum mechanical limit is $4\sqrt{2}$. Even
higher values (pattern-filled region) can be reached by allowing arbitrarily
strong nonlocal correlations.}
\end{center}
\end{figure}

\section*{Acknowledgments}
We acknowledge financial support from NSERC, OCE, and CFI.  We thank
$\rm{\check{C}}$aslav Brukner and Philip Walther for valuable
discussions. R.K. acknowledges financial support from IQC and ORDCF.
J.L. acknowledges financial support from the Bell Family Fund.

\end{document}